\documentclass[]{spie}  

\usepackage{amsmath,amsfonts,amssymb}
\usepackage{graphicx}
\usepackage[colorlinks=true, allcolors=blue]{hyperref}

\title{Simulating the performance of aperture mask designs for SCALES}

\author[a]{Mackenzie R. Lach}
\author[b]{Steph Sallum}
\author[c]{Andrew Skemer}
\affil[a]{University of California, Irvine, USA}
\affil[b]{University of California, Irvine, USA}
\affil[c]{University of California, Santa Cruz, USA}

\authorinfo{Further author information: (Send correspondence to M.R.L.)\\M.R.L.: E-mail: mrlach@uci.edu\\  S.S.: E-mail: ssallum@uci.edu}

\begin{document} 
\maketitle

\begin{abstract}
Interferometric techniques such as aperture masking have the potential to enhance spatial resolution capabilities when imaging moderate-contrast sources with small angular size, such as close-in exoplanets and circumstellar disks around distant young stars. The Slicer Combined with an Array of Lenslets for Exoplanet Spectroscopy (SCALES) instrument, currently under development, is a lenslet integral field spectrograph that will enable the W. M. Keck Observatory to carry out high-contrast direct imaging of exoplanets between 2 and 5 microns. We explore the potential benefit of aperture masking to SCALES by testing the contrast achievable by several mask designs. The \texttt{scalessim} software package was used to simulate observations at wavelength bins in the M, L, and K bands, with optical path difference (OPD) maps used to simulate realistic Keck adaptive optics performance. Noise from astrophysical and instrumental sources was also applied to simulated signals. Mask designs were assessed based on depth of the generated contrast curves.  
\end{abstract}


\section{INTRODUCTION}
\label{sec:intro}
The technique of high-contrast direct imaging enables the study and characterization of proto-planetary disks \cite{Esposito_2020, Hung_2015} and relatively young, hot planets that are at large angular separations from their stars. It can provide spectroscopic data on systems that are inaccessible to study via the transit or radial velocity methods, i.e. systems that do not transit or that have very wide angular separations \cite{Macintosh_2015}. A number of new exoplanets have been discovered and studied with this method \cite{Marois_2008, Haffert_2019}, and it has also allowed inferences to be made about the overall exoplanet population \cite{Bowler_2016}. The next generation of direct imaging instrumentation is expected to push detection limits and study a wider and more diverse population of planets, and will expand spectroscopic characterization capabilities.

SCALES, which stands for Slicer Combined with an Array of Lenslets for Exoplanet Spectroscopy, is an upcoming integral field spectrograph (IFS) that will operate from 2.0$\mu$m to 5.2$\mu$m and is expected to see first light at the W. M. Keck Observatory in 2025 \cite{Stelter_2020} (also see Ref. \citenum{Skemer_2022} in these Proceedings). As the first facility-class IFS that will operate at wavelengths longer than 2.5$\mu$m, SCALES will expand the currently accessible population of directly imaged planets to cooler and lower-mass planets. The high contrast between planet and host star is a major challenge for direct imaging; SCALES reduces this issue somewhat by operating in the thermal infrared, a wavelength regime where planet-star contrast is less extreme and where planets and disks are relatively bright. SCALES improves upon the design of the Arizona Lenslets for Exoplanet Spectroscopy (ALES) instrument, which is housed at the Large Binocular Telescope Interferometer (LBTI), and which provided the first spatially resolved spectra of substellar companions in the thermal infrared \cite{Skemer_2015}.

Non-redundant aperture masking (NRM) is a technique that is relatively cheap and simple to implement and that can significantly increase spatial resolution at moderate contrast (5-8 magnitudes at $\lambda/D$, depending on the image quality of the individual setup), surpassing the diffraction limit \cite{Tuthill_2000, Tuthill_2018, Sallum_2017, Sturmer_2012}. It consists of converting a telescope aperture into an interferometer by covering the aperture by an opaque mask with holes cut out. Typically this masking is done with a small mask in the instrumental pupil plane, often in a filter wheel. In a non-redundant mask, each pair of holes contributes uniquely to the Fourier transform of the fringes produced in the image plane, with no two baselines having the same length and angular position. In this work, we assess the performance of three different non-redundant aperture masks on SCALES. We utilize the Python-based \texttt{scalessim} software package\footnote{The \texttt{scalessim} package may be found at: \url{https://github.com/scalessim/scalessim}} to simulate SCALES observations using each mask. We then produce closure phases and squared visibilities and use the \texttt{SAMpy} package \cite{Sallum_2022} \footnote{\texttt{SAMpy} can be downloaded at: \url{https://github.com/JWST-ERS1386-AMI/SAMpy}} to generate contrast curves. 

\begin{figure} [ht]
\begin{center}
\begin{tabular}{c} 
\includegraphics[height=5.25cm]{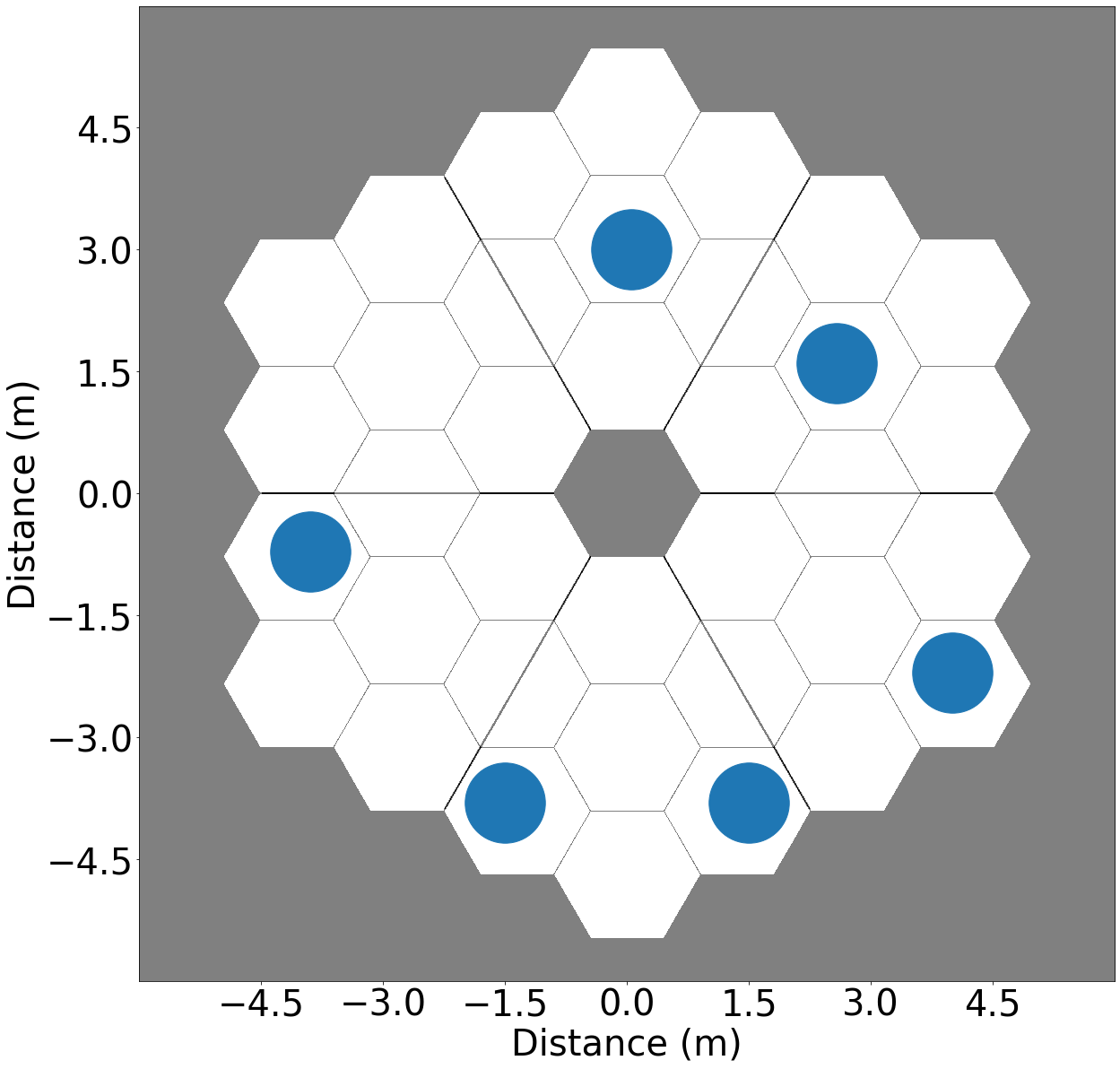}
\hfill
\includegraphics[height=5.25cm]{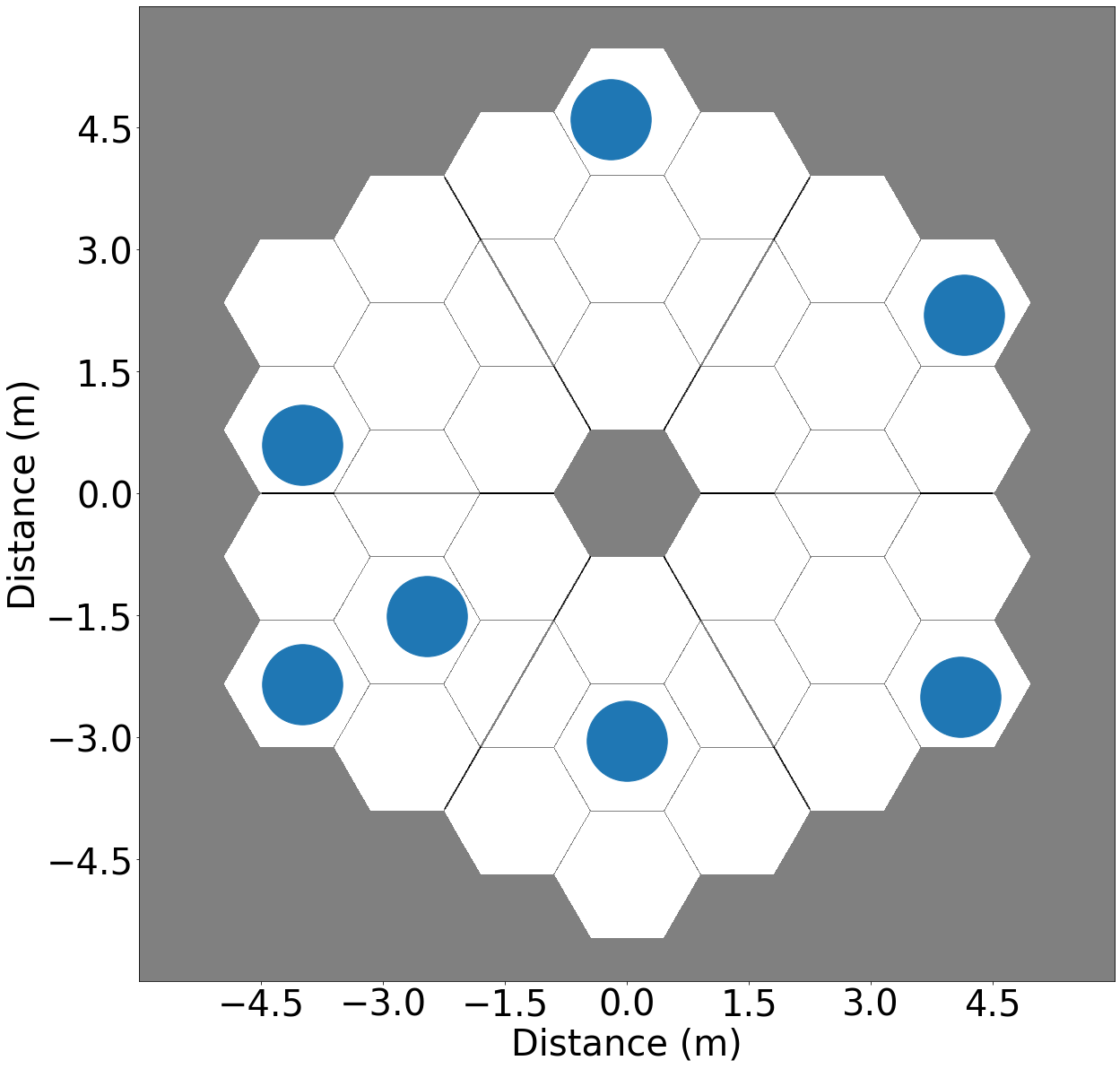}
\hfill
\includegraphics[height=5.25cm]{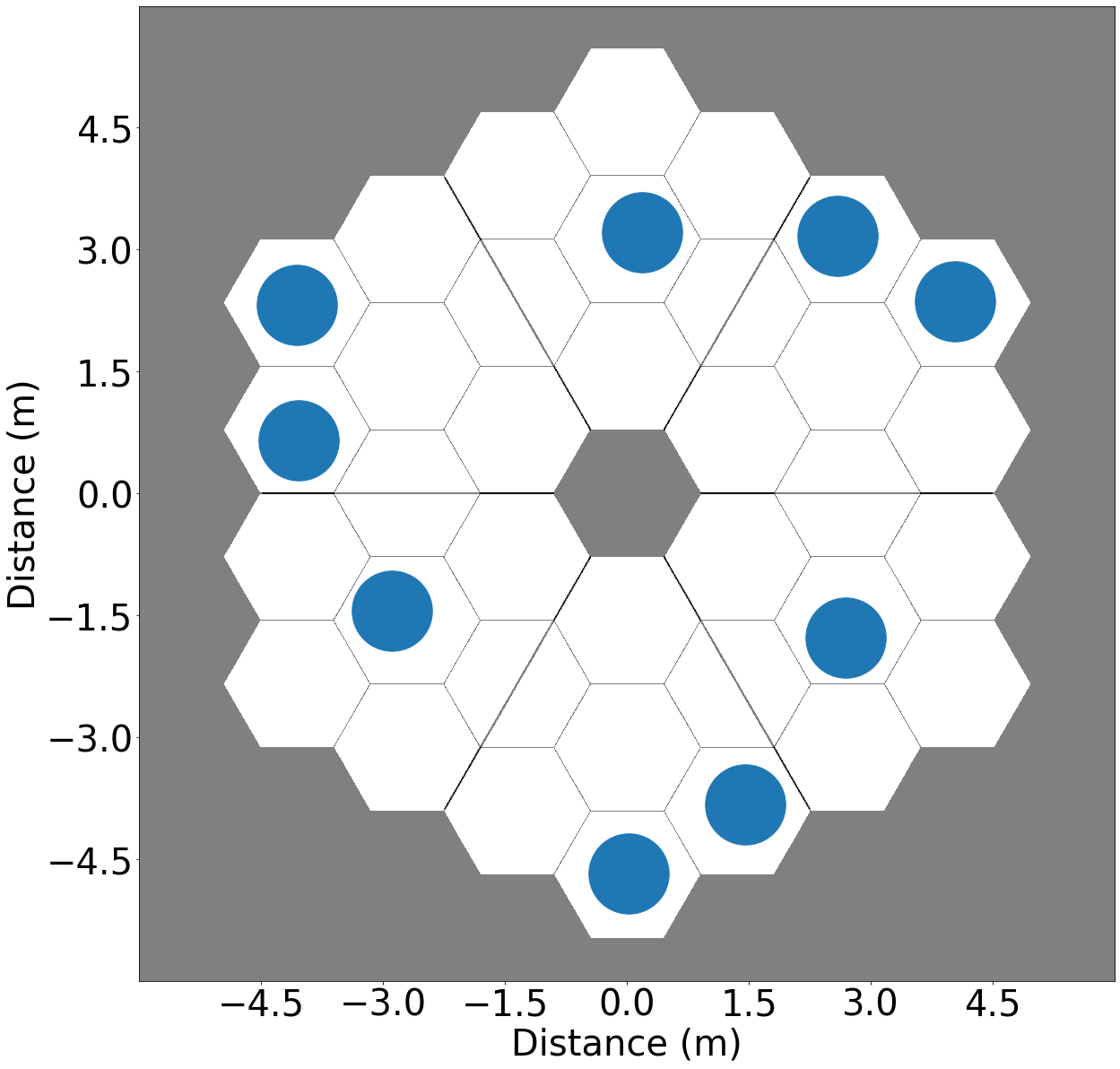}
\end{tabular}
\end{center}
\caption{The three non-redundant sparse aperture mask designs used in this work. The mask holes are shown projected onto the 10m Keck primary mirror and measure 1m in diameter. From left to right: the 6-hole mask modified from the LBT design; a 7-hole design modified slightly from the VLT SPHERE mask; the 9-hole mask currently on Keck/NIRC2.\label{fig:masks}}
\end{figure}

\section{Mask Designs}
\label{sec:procedures}
The design of sparse aperture masks that are fully non-redundant is challenging. Masks with large numbers of holes are difficult to produce, as are masks with large holes. This is due to the finite size of holes; a baseline between two holes can be chosen in more ways when the hole diameter is larger. Non-redundant masks with fewer or smaller holes are easier to design, although throughput is reduced. For these reasons, mask designs are frequently generated using computational methods \cite{Golay_1971, Tuthill_2000, Sturmer_2012}. 

Three mask designs were used in this study, with 9, 7, and 6 holes. These masks, as projected onto the Keck primary mirror, are shown in Fig. \ref{fig:masks}. To develop SCALES NRM simulation tools and as a first exploration of mask designs, the three masks were either chosen or modified from designs currently in use. The 9-hole mask design is currently in use in Keck/NIRC2. The 7- and 6-hole designs were modified from VLT SPHERE and LBT designs, respectively. The modifications were required in order to ensure that the holes do not overlap any mirror segment edges or spiders. Furthermore, several holes were adjusted in both in order to add longer baselines to the configuration and to ensure non-redundancy.

\section{Simulated Observations}
SCALES observations with each mask were simulated using the \texttt{scalessim} software. An observing scene was produced with a 10000K point source centered in the frame. A cube of PSFs was generated for each mask at oversampled spatial and spectral resolutions with respect to the SCALES low-resolution mode and was then convolved with the observing scene. The resulting cube was downsampled to coarser wavelength bins to match the SCALES low-spectral-resolution mode ($\lambda_{min} = 2.0\mu$m, $\lambda_{max} = 5.2\mu$m, $R<300$). A cube of raw images was then generated with an FOV of $2.2"$ x $2.2"$, spaxel plate scale of $0.02"$/pixel, and integration times of 1s and 3600s to simulate cases of short and long exposures. Ten frames were generated at each wavelength with random Poisson noise added. Further details about the scalessim data reduction pipeline are given in Ref.~\citenum{Breisemeister_2020}. \par
The Python-based \texttt{SAMpy} package was used to generate closure phases and squared visibilities. Closure phases for each mask, wavelength bin, and integration time simulated in this study are shown plotted versus average baseline length in Fig. \ref{fig:c_phases}. A supergaussian filter was applied to the SCALES frames to remove edge effects before these were calculated. The closure phases and squared visibilities were then calibrated as discussed in Section \ref{subsec:opd} and were used to generate contrast curves.
\label{sec:pipeline}

\subsection{OPD Maps and Calibration}
\label{subsec:opd}
Keck AO performance was modelled by applying realistic Keck AO residuals to the PSF cube. These data were simulated for the PyWFS operating in the H-band and with 0.66" seeing. One second of data were generated with a sampling rate of 1kHz. These data were averaged and then evolved using a degree 500 Zernike polynomial (see Ref. \citenum{Sallum_2019}) to represent changing wavefront error between a science target and a calibration PSF. Calibration PSFs are used to remove high-order noise that is not eliminated by closure phases or squared visibilities, which remove instrumental errors to first order \cite{Ireland_2013}. Care was taken to ensure that the standard deviation of the evolved OPD map matches that of the original OPD data. The evolved map reliably reproduces the image quality of Keck/NIRC2 with PyWFS. A background correction was performed by subtracting frames with no source injected. Further calibration was done by subtracting closure phases and dividing visibilities that were produced using the evolved OPD map.

\begin{figure} [ht]
\begin{center}
\begin{tabular}{c} 
\includegraphics[height=7cm]{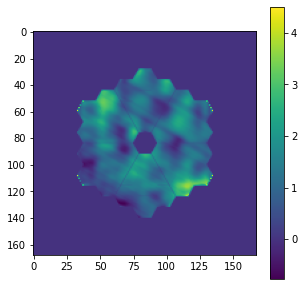}
\hfill
\includegraphics[height=7cm]{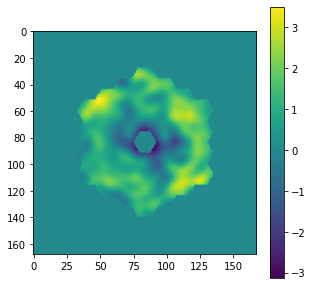}
\end{tabular}
\end{center}
\caption{OPD maps used in this work. Left: average of 1 second of data sampled at 1kHz. Right: averaged OPD map evolved using a 500 degree Zernike polynomial.\label{fig:opds}
}
\end{figure}

\section{Results}
\label{sec:results}

\begin{figure}[ht]
\begin{center}
\begin{tabular}{c} 
\includegraphics[height=5.5cm]{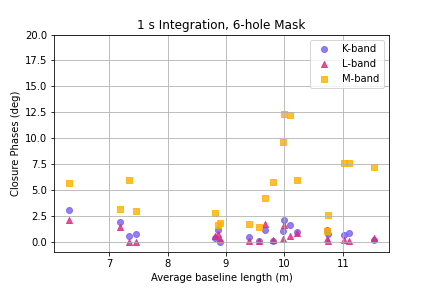}
\hfill
\includegraphics[height=5.5cm]{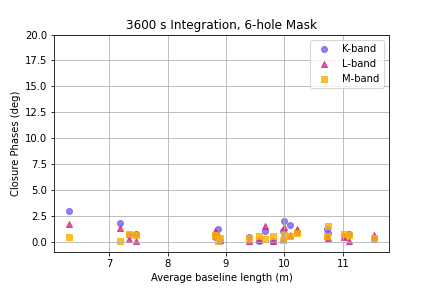}
\\
\includegraphics[height=5.5cm]{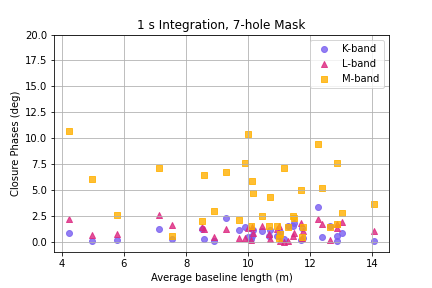}
\hfill
\includegraphics[height=5.5cm]{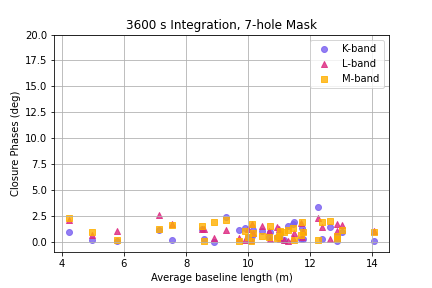}
\\
\includegraphics[height=5.5cm]{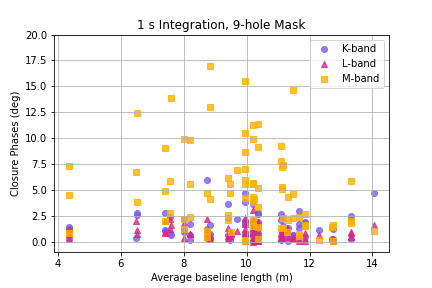}
\hfill
\includegraphics[height=5.5cm]{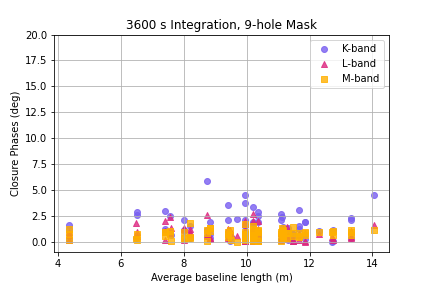}
\end{tabular}
\end{center}
\caption{Absolute values of closure phases for each mask at integration times of 1s and 3600s, plotted against the average baseline length calculated for each closure triangle. Data points are shown for individual wavelength bins in the K, L, and M bands.\label{fig:c_phases}
}
\end{figure}

\begin{figure}[ht]
\begin{center}
\begin{tabular}{c} 
\\
\includegraphics[height=5.25cm]{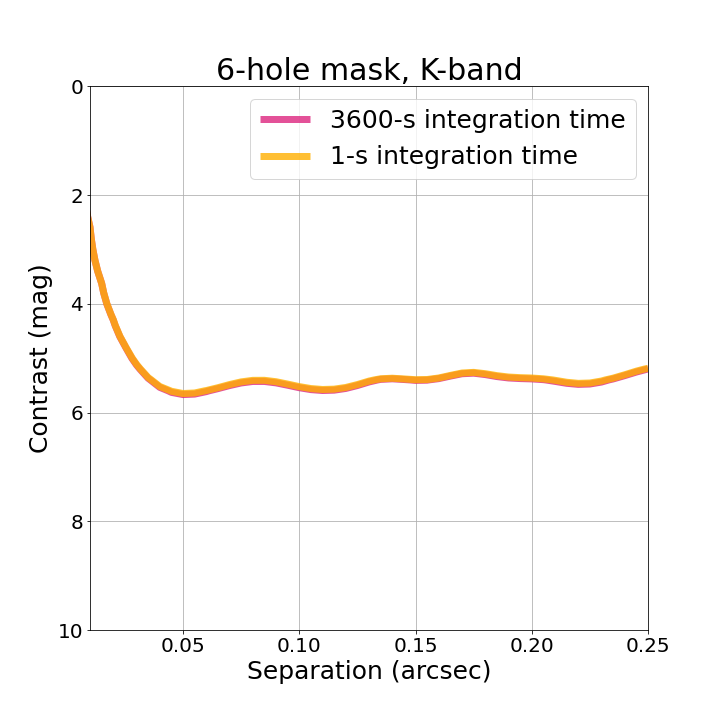}
\hfill
\includegraphics[height=5.25cm]{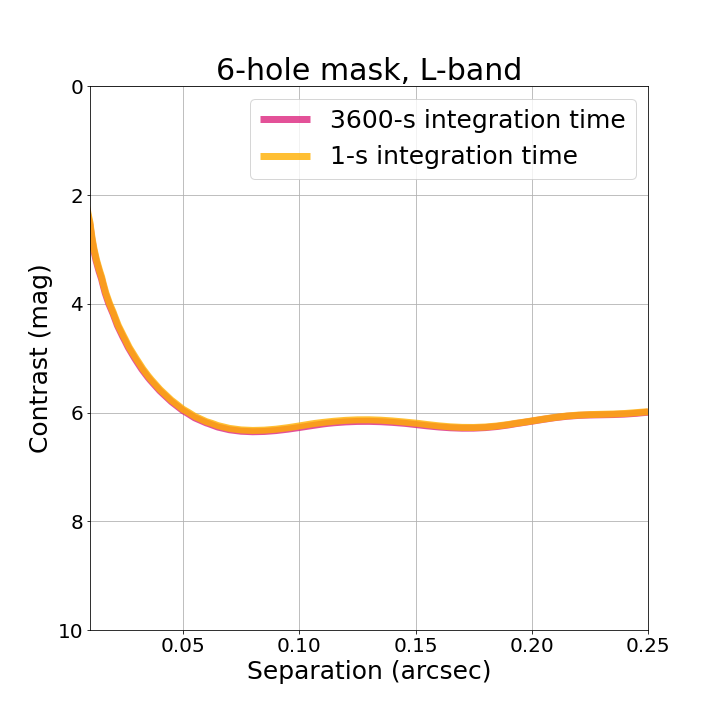}
\hfill
\includegraphics[height=5.25cm]{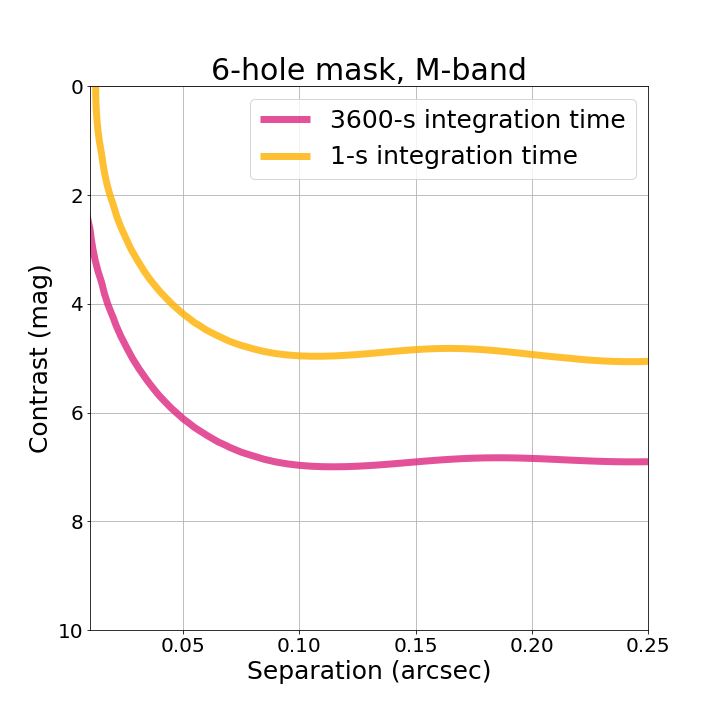}
\\
\includegraphics[height=5.25cm]{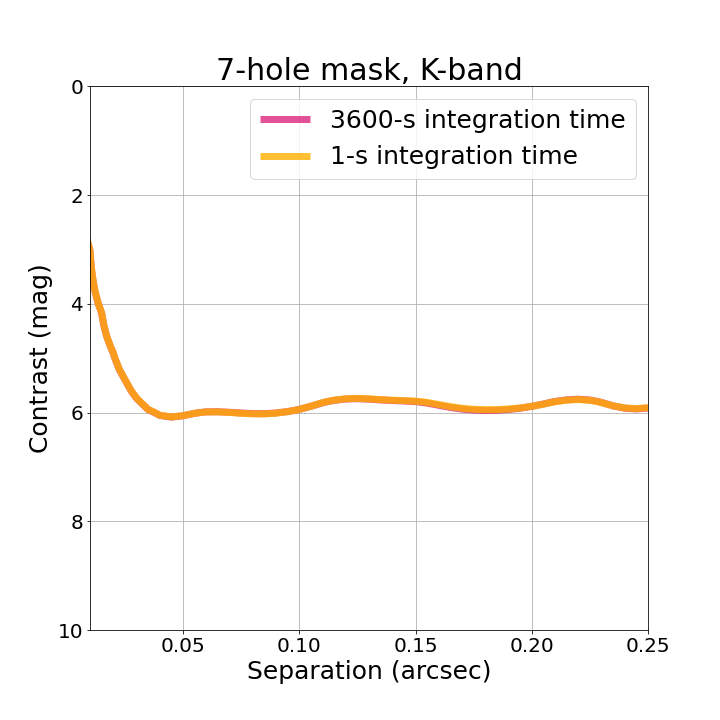}
\hfill
\includegraphics[height=5.25cm]{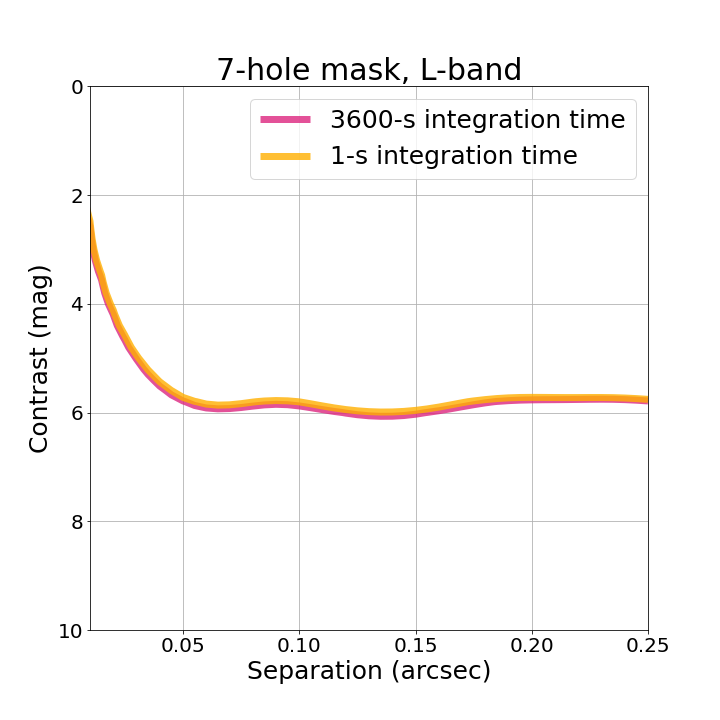}
\hfill
\includegraphics[height=5.25cm]{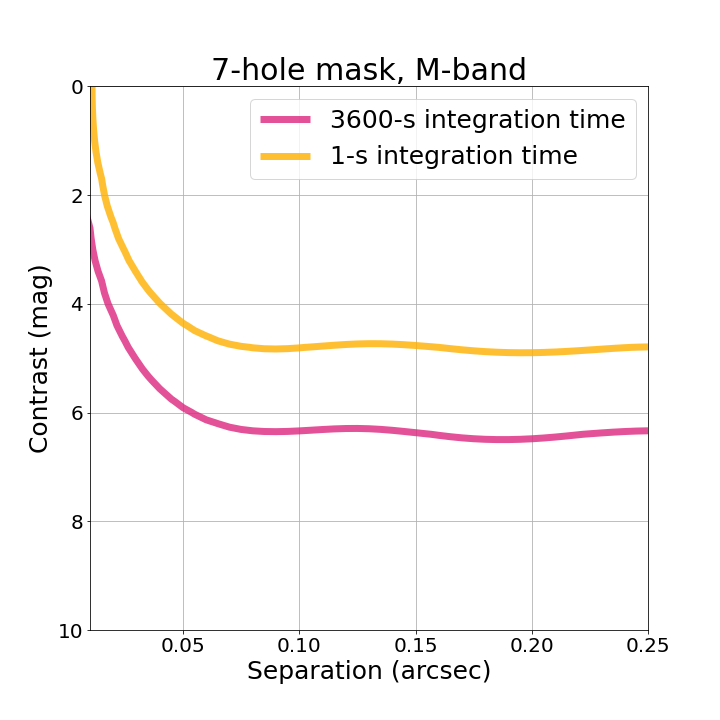}
\\
\includegraphics[height=5.25cm]{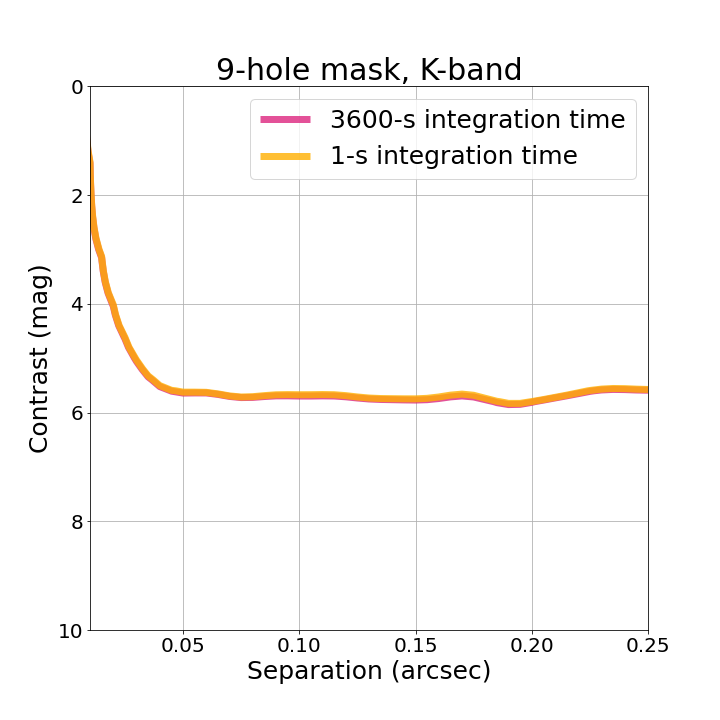}
\hfill
\includegraphics[height=5.25cm]{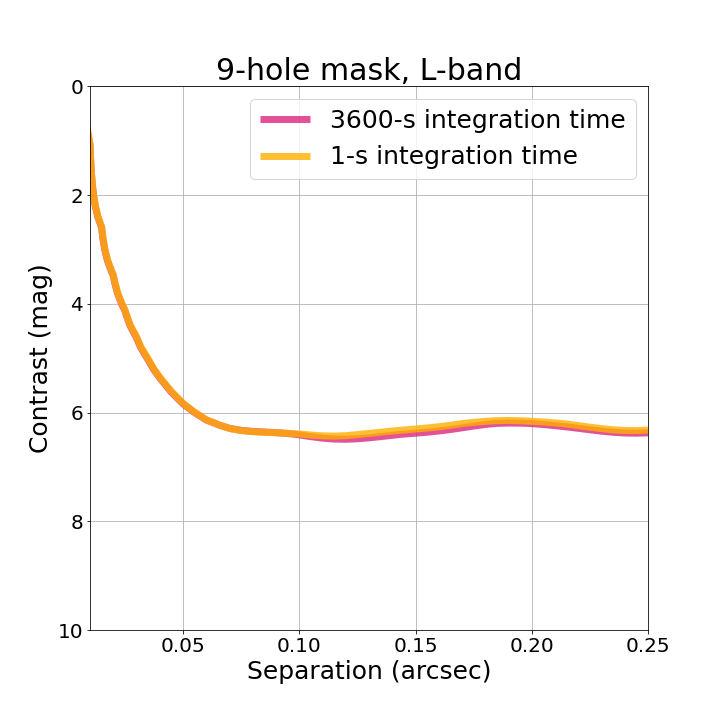}
\hfill
\includegraphics[height=5.25cm]{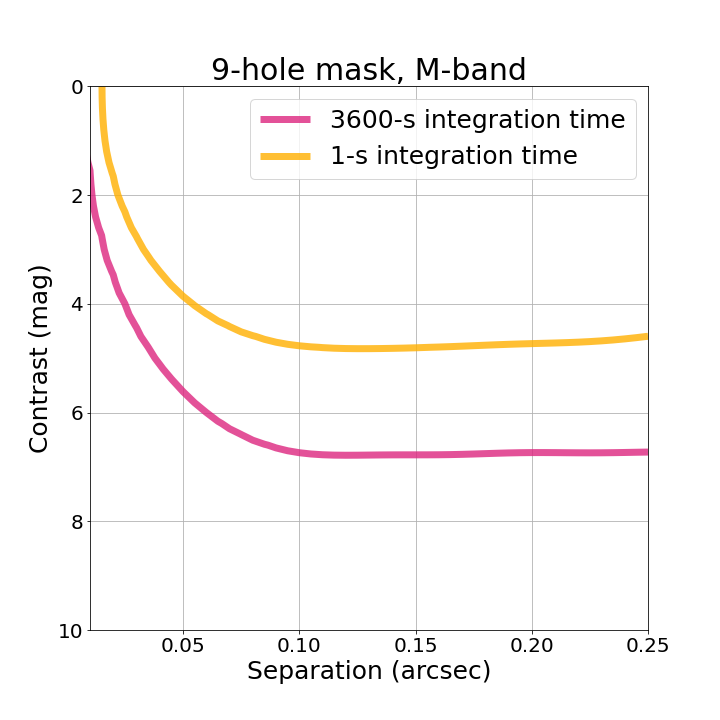}
\end{tabular}
\end{center}
\caption{$5-\sigma$ contrast curves are shown for each mask and at each wavelength bin used: $2.242\mu$m (K-band), $3.509\mu$m (L-band), and $5.019\mu$m (M-band).\label{fig:ccs}
}
\end{figure}

Fig. \ref{fig:c_phases} shows the calibrated absolute closure phases, for each mask and wavelength bin, for each closure triangle as a function of average baseline length for that triangle. Integration times of 1s and 3600s are shown. The contrast curves generated using these closure phases are shown in Fig. \ref{fig:ccs} for a $5-\sigma$ confidence interval. These are calculated by comparing $\chi^2$ values between single companion models to the null model for each dataset. Mask performance in a given band can be assessed by the depth of their contrast curves. In the K and L bands, a modest improvement in contrast can be observed between the 6- and 9-hole masks. The 3600s integration time does not achieve significantly better contrast than 1s in these bands due to the noise being dominated by OPD evolution as opposed to the sky, although there is some visible improvement with longer integration in the L-band. The M-band shows significant sky noise that is considerably mitigated with a longer integration. This can also be seen in the reduced closure phase scatter of Fig. \ref{fig:c_phases}. The 6-hole mask appears to offer improved contrast over the 7- and 9-hole masks in the background limit; this could be due to the simpler 6-hole pattern producing overall larger fringes that stand out better from the sky background. 

\section{Future Work}
Designing fully non-redundant aperture masks that are optimized for direct imaging studies is a challenging process. In future work, we will systematically generate a larger, more diverse sample of masks to test with the SCALES mask simulation pipeline presented here. Varying parameters such as number and size of holes, geometry, and degree of redundancy will be explored. Attempts will be made to achieve uniform u-v snapshot coverage that may be optimized for different targets. For example, masks with a wide range of baseline lengths, including several long baselines, could perform better when imaging extended targets with complex spatial features than masks lacking long baselines. Additional methods of assessing mask performance will also be implemented, including by injecting and attempting to recover a variety of science targets (e.g., companions, disks, or both). This work will be continued with the eventual goal of designing and testing a non-redundant aperture mask that is suitable for integration with SCALES.


\acknowledgments The authors would like to extend gratitude to Dr. Maaike van Kooten for providing the Keck OPD data utilized in this work. We would also like to thank the Heising-Simons Foundation for their support of the SCALES project.

\newpage

\newpage

\newpage

\bibliography{report} 
\bibliographystyle{spiebib} 

\end{document}